\documentclass[submission,copyright,creativecommons]{eptcs}
\providecommand{\event}{WORDS 2011} 
\usepackage{breakurl}             
\usepackage{amssymb}

\title{Finite-Repetition threshold\\for infinite ternary words}
\author{Golnaz Badkobeh
\institute{King's College London, UK}
\email{golnaz.badkobeh@kcl.ac.uk}
\and
 Maxime Crochemore
\institute{King's College London, UK}
\institute{Universit\'e Paris-Est, France}
\email{maxime.crochemore@kcl.ac.uk}
}
\def\titlerunning{Finite-Repetition threshold for infinite ternary words}
\def\authorrunning{G. Badkobeh and M. Crochemore}

\newcommand{\dd}{\mathinner{\ldotp\ldotp}}   
\newtheorem{theorem}{Theorem}

\newtheorem{proposition}{Proposition}
\newtheorem{conjecture}{Conjecture}
\newtheorem{lemma}{Lemma}

\newcommand{\per}{\textit{period}}

\newcommand{\FRt}{\textrm{FRt}}

\begin{document}
\maketitle

\begin{abstract}

The exponent of a word is the ratio of its length over its smallest period.
The repetitive threshold $r(a)$ of an $a$-letter alphabet is the smallest
 rational number for which there exists an infinite word whose finite factors
 have exponent at most $r(a)$.
This notion was introduced in 1972 by Dejean who gave the exact values of $r(a)$
 for every alphabet size $a$ as it has been eventually proved in 2009.

The finite-repetition threshold for an $a$-letter alphabet refines the above notion.
It is the smallest rational number $\FRt(a)$ for which there exists an
 infinite word whose finite factors have exponent at most $\FRt(a)$
 and that contains a finite number of factors with exponent $r(a)$.
It is known from Shallit (2008) that $\FRt(2)=7/3$.

With each finite-repetition threshold is associated the smallest number of
 $r(a)$-exponent factors that can be found in the corresponding infinite word.
It has been proved by Badkobeh and Crochemore (2010) that this number is $12$ for
 infinite binary words whose maximal exponent is $7/3$.

We show that $\FRt(3)=r(3)=7/4$ and that the bound is achieved with
 an infinite word containing only two $7/4$-exponent words, the smallest number.

Based on deep experiments we conjecture that $\FRt(4)=r(4)=7/5$.
The question remains open for alphabets with more than four letters.


\end{abstract}

\section {Introduction}

The article deals with repetitions in strings and their avoidability.
The question is grounded on the notion of the exponent of a word: it is
 the ratio of its length over its smallest period.
A word of exponent $e$ is also called an $e$-power.

An infinite word is said to avoid $e$-powers or to be $e$-power free
 if the exponents of its finite factors are smaller than $e$.

The repetitive threshold $r(a)$ of an $a$-letter alphabet is the smallest
 rational number for which there exists an infinite word whose finite factors
 have exponent at most $r(a)$.
The word is said to be $r(a)^+$-power free.
It is known from Thue \cite{Thue06} that $r(2)=2$.
Indeed, the notion was introduced in 1972 by Dejean \cite{Dej72} who proved
 that $r(3)=7/4$ and gave the exact values of $r(a)$
 for every alphabet size $a>3$.
Her conjecture was eventually proved in 2009 after partial proofs given by several
 authors (see \cite{Rao11,CR11} and references therein).

A generalised version of the repetitive threshold is by Ilie et al. \cite{IlieOS05}.
The authors introduce the notion of $(\beta, p)$-freeness:
 a word is $(\beta, p)$-free if it contains no factor that is a
 $(\beta', p')$-repetition (i.e. a word of period $p'$ and exponent $\beta'$)
 for $\beta'\geq \beta$ and $p'\geq p$;
 it is $(\beta^+, p)$-free if $\beta'> \beta$ instead.
Their generalized repetition threshold $R(a,p)$ defined for an $a$-letter alphabet
 as the real number $\alpha$ for which either
\begin{itemize}\itemsep0mm\topsep0mm
\item[(a)]
 there exists an $(\alpha^+,p)$-free infinite
 word and all $(\alpha,p)$-free words are finite,
\item[(b)]
 or there exists an $(\alpha,p)$-free infinite word and
 for all $\epsilon > 0$,  $(\alpha-\epsilon,p)$-free words are finite.
\end{itemize}
 where $p$ is the minimum avoided period.
The proof of boundary of this threshold for all alphabet sizes is presented in \cite{IlieOS05},
 and $R(k,1)$ is essentially Dejean's repetition threshold.


For infinite words whose maximal exponent of factors is bounded, it is legitimate
 to ask whether they can contain only a finite number of $r(a)$-powers.
This is an extra constraint on the word.
When such words exist, it is as legitimate to exhibit the minimal number of
 $r(a)$-powers they can contain, which adds another measure of the word complexity.
The first result of this type is by Fraenkel and Simpson \cite{FS95} for the
 binary alphabet.
They showed that an infinite binary word can contain only 3 squares, not less.
Two simple proofs of the result are by Harju and Nowotka \cite{HN06} and the present
 authors \cite{BC10-3sq}.

The above consideration leads to the notion of finite-repetition threshold associated
 with an $a$-letter alphabet.
It is the smallest rational number, noted $\FRt(a)$, for which there exists an
 infinite word whose finite factors have exponent at most $\FRt(a)$
 and that contains a finite number of $r(a)$-powers.
It is known from Shallit \cite{Sha04} that $\FRt(2)=7/3$ (see also \cite{RampersadSW05}).
The present authors \cite{BC1012sq-rairo} proved
 that the associated minimal number of squares is $12$ if the infinite word contains two $7/3$-powers.
Badkobeh \cite{Bad11} even refined the results by showing the number is $8$ if the infinite word
 admits two $5/2$-powers, extending the result of Fraenkel and Simpson \cite{FS95}
 recalled above for which it is $3$ if two cubes are allowed.

In this article, we consider the finite-repetition threshold of the ternary alphabet.
We show that $\FRt(3)=r(3)=7/4$.
We provide a direct proof of the result and another proof
 based on a previous result on word morphisms by Ochem \cite{Ochem06}.

The experiments reported in the conclusion show that the finite-repetition threshold
 of the $4$-letter alphabet $\FRt(4)$ is likely to be $r(4)=7/5$, which we conjecture.
The hypothetical property $\FRt(a)=r(a)$ (for $a>2$) would be equivalent to say that
 infinite words whose maximal exponent of factors is Dejean's repetition threshold
 can be constrained to containing a finite number of factors with that exponent.

\section{Repetitions in ternary words}

Let $A$ be a finite alphabet.
A word $w$ in $A^*$ of length $|w|=n$ is a sequence of letters $w[0]w[1]\dots w[n-1]$
 also noted $w[0\dd n-1]$.
The period of $w$ is the smallest positive integer $\per(w)=p$ for which $w[i]=w[i+p]$ whenever
 both sides of the equality are defined.
The exponent of $w$ is the rational ratio $|w|/\per(w)$.
Thus, the exponent of a word is a rational number that is at least $1$.
For example, a square is a nonempty word with an even integer exponent
 and $\mathtt{1020102}$ of exponent $7/4$ can be written $(\mathtt{1020})^{7/4}$.
A word of exponent $e$ is also called an $e$-power.

An infinite word is a function from the natural number to the alphabet $A$.
An infinite word is said to avoid $e$-powers (resp. $e^+$-powers) if the exponents of
 its finite factors are smaller than $e$ (resp. not more than $e$).
In this case we also say that the word is $e$-power free (resp. $e^+$-power free).

The repetitive threshold $r(a)$ of an $a$-letter alphabet is the smallest
 rational number for which there exists an infinite word whose finite factors
 have exponent at most $r(a)$.
The word is then $r(a)^+$-power free.

The \textit{finite-repetition threshold} for the alphabet of $a$ letters
 is defined as the smallest rational number $\FRt(a)$ for which
 there exists an infinite word that both avoids $\FRt(a)^+$-powers
 and contains a finite number of $r$-powers, where $r$ is Dejean's repetitive threshold.

The above notion is inspired by the following results of Karhum\"aki and Shallit.
 \cite{Sha04}
\begin{theorem}[Karhum\"aki and Shallit \cite{KarhumakiS04}]
For all $t \geq 1$, there are no infinite binary words that simultaneously avoid all squares
 $yy$ with $|y| \geq t$ and $7/3$-powers.
\end{theorem}

\begin{theorem}[Shallit \cite{Sha04}]
There is an infinite binary word that simultaneously avoids all squares $yy$ with $|y|\geq 7$
 and ${7/3}^+$-powers.
\end{theorem}

When an infinite binary word avoids ${7/3}^+$-powers and contains a finite number of
 squares it is natural to ask more on these few squares.
The previous theorem shows that their period can be bounded by $7$.
The next result goes slightly beyond by showing that their number is at least $12$.
But the two properties cannot be satisfied simultaneously.

\begin{theorem}[Crochemore and Badkobeh \cite{BC1012sq-rairo}]\label{theo-1}
The smallest number of squ\-ares occurring in a $7/3$-power free infinite binary word
 is $12$.
\end{theorem}

Showing that no infinite word satisfying the conditions can contain less squares is done
 by mere computation.
The second part is done by producing an infinite word satisfying the condition
 and containing exactly $12$ squares.
Following Shallit's hierarchy of infinite binary words in \cite{Sha04}, the previous result
 was refined by Badkobeh \cite{Bad11} according to the next table.

\begin{center}
\begin{tabular}{|c|c|c|}
      \hline
     Maximal      & Allowed number   & Smallest number \\
     exponent $e$ & of $e$-powers    & of squares     \\
      \hline
     7/3              & 2                       & 12   \\
                      & 1                       & 14  \\
      \hline
     5/2              & 2                       & 8   \\
                      & 1                       & 11   \\
      \hline
     3                & 2                       & 3   \\
                      & 1                       & 4   \\
       \hline
\end{tabular}
\end{center}

The main result of the present article is the following theorem.

\begin{theorem}\label{theo-2}
The \textit{finite-repetition threshold} of the 3-letter alphabet is its Dejean's repetition
 threshold, that is, $7/4$.\\
The smallest number of $7/4$-powers occurring in a ${7/4}^+$-power free
 infinite ternary word is $2$.
\end{theorem}

On the alphabet $\{\mathtt{0}, \mathtt{1}, \mathtt{2}\}$, the two unavoidable $7/4$-powers
 occurring in the word below are, up to a permutation of letters,
 $(\mathtt{0121})^{7/4} = \mathtt{0121012}$ and $(\mathtt{2010})^{7/4} = \mathtt{2010201}$.

Computation shows that the longest ternary words with only one $7/4$-power
 are $102$ letters long.
However we may think of having a larger threshold as in the binary case.
But even if we increase the threshold to $e<2$, the maximal length of words
 stays at $102$ with only one $e$-power.

If we relax further the maximal exponent condition, it can be shown that there exists
 an infinite ternary word in which occur only one square, namely $\texttt{00}$ up to a
 permutation of letters, and no $e$-power with $7/4 \leq e <2$.

%

Since the repetition threshold for a 3-letter alphabet is $7/4$, to prove this ratio is
 also its finite-repetition threshold it is sufficient to show (contrary to the binary case)
 that there exists a ${7/4}^+$-free infinite ternary word with finitely many $7/4$-powers.
To do it, we use the fact that the repetition threshold of $4$-letter alphabets is $7/5$
 and provide a translation morphism from $4$ letters to $3$ letters with suitable conditions.

We consider the morphism $g$ from $\{\mathtt{a}, \mathtt{b}, \mathtt{c}, \mathtt{d}\}^*$ to
 $\{\mathtt{0}, \mathtt{1}, \mathtt{2}\}^*$ defined by:
$$\cases{
  g(\mathtt{a}) = & $\mathtt{0102101202102010210121020120210120102120121020120210121}$ \cr
& $\mathtt{0212010210121020102120121020120210121020102101202102012}$ \cr
& $\mathtt{10212010210121020120210120102120121020102101210212}$, \cr
  g(\mathtt{b}) = & $\mathtt{0102101202102010210121020120210120102120121020120210121}$ \cr
& $\mathtt{0201021012021020121021201021012102010212012102012021012}$ \cr
& $\mathtt{10212010210121020120210120102120121020102101210212}$, \cr
  g(\mathtt{c}) = & $\mathtt{0102101202102010210121020120210120102120121020102101202}$ \cr
& $\mathtt{1020121021201021012102010212012102012021012102010210120}$ \cr
& $\mathtt{21020102120121020120210120102120121020102101210212}$,\cr
  g(\mathtt{d}) = & $\mathtt{0102101202102010210121020120210120102120121020102101202}$ \cr
& $\mathtt{1020102120121020120210121020102101202102012102120102101}$ \cr
& $\mathtt{21020102120121020120210120102120121020102101210212}$.
}$$
The morphism is uniform with codeword length $160$.
Another presentation of the morphism $g$ is:
$$\cases{
g(\mathtt{a}) = u v \mathtt{02120121020120210121020102101202102012102120102101} y z, & \cr
g(\mathtt{b}) = u v \mathtt{21021201021012102010212012102012021012102010210120} y z, & \cr
g(\mathtt{c}) = u w \mathtt{01021012021020121021201021012102010212012102012021}x z, &\cr
g(\mathtt{d}) = u w \mathtt{12010210121020102120121020120210121020102101202102} x z,
}$$
where $u$, $v$, $w$, $x$, $y$ and $z$ are:\\
$u = \mathtt{01021012021020102101210201202101201021201210201}$,\\
$v = \mathtt{2021012102}$,
$w = \mathtt{0210120210201}$,
$x = \mathtt{2102010212}$,
$y = \mathtt{0121021201021}$,\\
$z = \mathtt{0121020120210120102120121020102101210212}$.\\
The word $u$ is the longest common prefix of the codewords, $|u|=47$,
 and $z$ is their longest common suffix, $|z|=40$.

Theorem~\ref{theo-2} is a direct consequence of the next proposition.

\begin{proposition}\label{pro-1}
The morphism $g$ translates any infinite ${7/5}^+$-free word 
 on the alphabet $\{\mathtt{a}, \mathtt{b}, \mathtt{c}, \mathtt{d}\}$
 into a ${7/4}^+$-free ternary word containing only two $7/4$-powers, the fewest possible.
\end{proposition}

We present two proofs of Proposition \ref{pro-1}.
The first one is a direct proof that involves the longest common prefix
 and the longest common suffix of the codewords to derive a contradiction
 from the existence of any $7/4$-power other than $\mathtt{0121012}$ and
 $\mathtt{2010201}$ in the image by $g$ of a ${7/5}^+$-free word.
The second proof is derived from a lemma on morphisms stated by Ochem in \cite{Ochem06}.

\subsection*{Direct proof of Proposition \ref{pro-1}}

Let us assume that $g(s)$ contains a non-extensible repetition, excluding
 the two $7/4$-powers $\mathtt{0121012}$ and $\mathtt{2010201}$,
 with exponent at least $7/4$.
The repetition can be written $pq$ where $|p|$ is its period.
Then $|pq|/|p| \geq 7/4$.
A simple computation verifies that no image of a ${7/5}^+$-free word with length at most $3$
 contains the repetition.
Therefore the repetition is long and occurs in the image by $g$ of a word of length at least $4$.

We consider two cases.
\begin{itemize}
 \item Case $|p|\leq |q|$.
The word $pq$ is of the form
$$pq= 
 \overbrace{u_1\underbrace{ \cdots }v_1}\overbrace{u_1 \underbrace{ \cdots } v_1} \cdots $$
where $u_1v_1$ is codeword.
Indeed it starts with the square $pp$ of the form
$$\overbrace{u_1 g(s')v_1}\overbrace{u_1 g(s') v_1}$$ 
where $s' \in \{\mathtt{a}, \mathtt{b}, \mathtt{c}, \mathtt{d}\}^*$.

Note that $s'$ cannot be the empty word because $pq$ would occur in the image of a triplet.

Let $\alpha \in \{a, b, c, d\}$ be such that $g(\alpha) = v_1 u_1$.
Therefore $s' \alpha s'$ is a factor of $s$.
The letter occurring before $s'$ in $s$ and the letter occurring after it
 must differ from $\alpha$ to avoid the squares $\alpha s' \alpha s'$ or
 $s' \alpha s' \alpha$ since $s$ is ${7/5}^+$-power free).

Then $u_1$ is not longer than the longest common prefix between two different codewords,
 that is, $|u_1| \leq |uw| = 60$.
Symmetrically, $v_1$ is not longer than the longest common suffix of two different
 codewords, that is, $|v_1| \leq |yz| = 53$.
But then $|v_1u_1| \leq 113$ and cannot be a complete codeword, a contradiction.

\item Case $|p|> |q|$.
The word $pq$ is of the form
$$pq = \overbrace{u_1\underbrace{ \cdots }v_1}\cdots \overbrace{u_1 \underbrace{ \cdots } v_1} $$
More precisely $a_0pqb_1$ is of the form
$$a_0\underbrace{ u_1g(s') v_1}a_1\cdots  b_0 \underbrace{ u_1 g(s') v_1 }b_1$$
 where $s' \in \{\mathtt{a}, \mathtt{b}, \mathtt{c}, \mathtt{d}\}^*$,
 $a_0, a_1, b_0, b_1 \in \{\mathtt{0}, \mathtt{1}, \mathtt{2}\}$ and
 $a_0\neq b_0$ and $a_1\neq b_1$, because $pq$ is inextensible.
It rewrites as
$$a_0 u_1 g(s') g(s'') g(s') v_1 b_1$$
 where $g(s'')=v_1 q^{-1}p u_1$ because the morphism is synchronizing (no codeword occurs
 in the concatenation of two codewords).
Therefore $g(s')g(s'')g(s')$ is a factor of $g(s)$
 thus $s's''s'$ is a factor of $s$ and since $s$ is  ${7/5}^+$-free we get
$$\frac{|s's''s'|}{|s's''|} \leq \frac{7}{5}$$
and
$$ 3|s'| \leq 2|s''| 
$$ 
and eventually
\begin{eqnarray}
3|g(s')| \leq 2|g(s'')|\label{eq:2}
\end{eqnarray}
 because the morphism $g$ is uniform.

Furthermore $pq = u_1 g(s') g(s'') g(s') v_1$ and $p= u_1 g(s') g(s''){u_1}^{-1}$
 so its exponent 
 satisfies
\begin{eqnarray*}
\frac{|u_1 g(s') g(s'') g(s') v_1|}{|g(s') g(s'')|} &\geq& \frac{7}{4}
\end{eqnarray*}
 which rewrites as
\begin{eqnarray}
|g(s')|+ 4|u_1v_1| &\ge & 3|g(s'')| \label{eq:3}
\end{eqnarray}

Using Equations \ref{eq:2} and \ref{eq:3} we get
\begin{eqnarray*}
 9|g(s')| & \leq & 6|g(s'')| \cr
          & \leq & 2(|g(s')|+ 4|u_1v_1|)
\end{eqnarray*}
and then
\begin{eqnarray*}
 |g(s')| & \leq & \frac{8}{7}|u_1v_1|.
\end{eqnarray*}
But since $|u_1v_1|\leq 113$ as in the first case, this implies that $s'$ is empty.
Therefore the repetition $pq$ is a factor of the image of a triplet,
 a contradiction.
\end{itemize}
This completes the direct proof of Proposition \ref{pro-1}.

\subsection*{Proof based on Ochem's result}

Here we split the proof in two parts: first we show that $g(s)$ is ${7/4}^+$-free,
 second we show the only $7/4$-powers are the ones mentioned above.
The proof depends on the following result.
In the statement, $\Sigma_s$ (resp. $\Sigma_e$) is an alphabet with $s$ (resp. $e$) letters;
 and the morphism $h:\Sigma^*_s\rightarrow \Sigma^*_e$ is synchronizing if
 for any $a, b, c \in \Sigma_s$ and $v, w \in \Sigma^*_e$,
 $h(ab) = vh(c)w$ implies either $v=\epsilon$ and $a = c$ or $w = \epsilon$ and $b = c$.

\begin{lemma}[Ochem \cite{Ochem06}]\label{lemma-1}
Let $\alpha, \beta \in \mathbb{Q}$, $1<\alpha<\beta<2$, and $p \in \mathbb{N}^*$.
Let $h:\Sigma^*_s\rightarrow \Sigma^*_e$ be a synchronizing $q$-uniform morphism (with $q \geq 1$).
If $h(w)$ is ($\beta^+, p$)-free for every $\alpha^+$-free word $w$ such that
 $|w| < \max\{\frac{2\beta}{\beta-\alpha}, \frac{2(q-1)(2\beta-1)}{q(\beta-1)} \}$,
 then $h(t)$ is $(\beta^+, p)$-free for every (finite or infinite)
 $\alpha^+$-free word $t$.
\end{lemma}

To apply the lemma to the morphism $g$ above, we have $\alpha=7/5$ and $q=160$.
We choose $\beta= 17/10$ and $p=5$.
Then we can show that the morphism is $({17/10}^+, 5)$-free if $g(w)$ is $({17/10}^+,5)$-power free
 for all words $w$ for which
 $$|w| < \max\{\frac{2\beta}{\beta-\alpha}, \frac{2(q-1)(2\beta-1)}{q(\beta-1)} \},$$
 which implies $|w| < 12$.
This set is finite and a simple computation can verify the claim.

Since every $({7/4}^+, 5)$-power is also a $({17/10}^+, 5)$-power then we can claim
 the morphism is $({7/4}^+, 5)$-power free.

So the only possible $7/4$-powers with period less that $5$ are:\\
 $(0121)^{7/4}$, $(0212)^{7/4}$, $(1020)^{7/4}$, $(1202)^{7/4}$, $(2010)^{7/4}$, and $(2101)^{7/4}$.
Any of those strings must be either a factor of a codeword or a factor of the image of a doublet.
We immediately conclude the words
 $(\mathtt{1020})^{7/4}$ and $(\mathtt{2101})^{7/4}$ are the only factors of $g(s)$.

This concludes the whole proof of Theorem \ref{theo-2}.

\section{Repetitions for larger alphabets}

Experiments show that a word on a $4$-letter alphabet for which the maximal exponent of factors is $7/5$
 and that contains at most one $7/5$-power has maximal length $230$.
However if the constraint on the number of $7/5$-powers is relaxed to $2$
 the length grows to at least $100000$.
This experiment intrigued us to study the string further and to state the following conjecture.

\begin{conjecture}
The \textit{finite-repetition threshold} of 4-letter alphabets is $7/5$
 and their exist an infinite ${7/5}^+$-power free word containing only two $7/5$-powers.\\
The two $7/5$-powers are $(0231203213)^\frac{7}{5}$ and $(1230213203)^\frac{7}{5}$
 up to a permutation of the letters.
\end{conjecture}

Note the above words have period $10$, the smallest possible period, since there is
 no $7/5$-power with period $5$ that is ${7/5}^+$-free.

The experiments are done with a mere backtracking technique to generate the suitable words.
It implements efficient algorithms for testing the properties.

\section*{Acknowledgments}

We warmly thank Pascal Ochem for pointing his powerful result on word morphisms
 and for profitable discussions.

\bibliographystyle{eptcs}
\bibliography{ternary}

\end{document}